\documentclass[10pt,twocolumn,letterpaper]{article}

\usepackage[pagenumbers]{wacv} 

\usepackage{graphicx}
\usepackage{amsmath}
\usepackage{amssymb}
\usepackage{booktabs}
\usepackage{adjustbox}
\usepackage{multirow}
\usepackage{tcolorbox}
\usepackage{pythonhighlight}
\usepackage{makecell}

\usepackage{pifont}
\usepackage[pagebackref,breaklinks,colorlinks]{hyperref}

\usepackage[capitalize]{cleveref}
\crefname{section}{Sec.}{Secs.}
\Crefname{section}{Section}{Sections}
\Crefname{table}{Table}{Tables}
\crefname{table}{Tab.}{Tabs.}

\def\wacvPaperID{*****} 
\def\confName{WACV}
\def\confYear{2025}

\begin{document}

\title{LiteGPT: Large Vision-Language Model for Joint Chest X-ray \\Localization and Classification Task}

\author{Khai Le-Duc$^{*1,3,4}$, Ryan Zhang$^{*1,2,3}$, Ngoc Son Nguyen$^{*4,5}$, \\Tan-Hanh Pham$^{6}$, Anh Dao$^{7}$, Ba Hung Ngo$^{8}$, Anh Totti Nguyen$^{9}$, Truong-Son Hy$^{4,10}$ \\
\hphantom{text}\\
$^1$University of Toronto, Canada 
$^2$Johns Hopkins University, USA\\
$^3$University Health Network, Canada 
$^4$FPT Software AI Center, Vietnam\\
$^5$VNUHCM - University of Science, Vietnam 
$^6$Florida Institute of Technology, USA\\
$^7$Michigan State University, USA
$^8$Chonnam National University, South Korea\\
$^9$Auburn University, USA 
$^{10}$Indiana State University, USA \\
{\tt\small $^1$duckhai.le@mail.utoronto.ca, $^{10}$TruongSon.Hy@indstate.edu}
}
\maketitle

\begin{abstract}
Vision-language models have been extensively explored across a wide range of tasks, achieving satisfactory performance; however, their application in medical imaging remains underexplored. In this work, we propose a unified framework - LiteGPT - for the medical imaging. We leverage multiple pre-trained visual encoders to enrich information and enhance the performance of vision-language models. To the best of our knowledge, this is the first study to utilize vision-language models for the novel task of joint localization and classification in medical images. Besides, we are pioneers in providing baselines for disease localization in chest X-rays. Finally, we set new state-of-the-art performance in the image classification task on the well-benchmarked VinDr-CXR dataset. All code and models are publicly available online.
\end{abstract}

\def\thefootnote{*}\footnotetext{Equal contribution}\def\thefootnote{\arabic{footnote}}

\section{Introduction}
\label{sec:intro}
The advent of large-scale vision-language models (VLMs) has transformed the fields of computer vision and natural language processing with their versatility and interactivity, surpassing the capabilities of vision-only models \cite{achiam2023gpt4, team2023gemini}. VLMs, trained on extensive datasets of image-text pairs, excel in various tasks, including image classification \cite{he2017fine}, image captioning \cite{hu2022scaling}, and visual question answering (VQA) \cite{seenivasan2023surgicalgpt}. They enable more interactive human-AI applications, such as conversational AI that can discuss visual content \cite{wang2023enabling}, provide explanations \cite{sammani2022nlx}, and more.

Despite these advancements, the application of VLMs to the medical domain remains under-explored. Recently, the topic of VLMs for the medical domain has gained significant interest among researchers. This is because their applications in real-life medical settings can be highly beneficial for doctors, automating routine tasks and assisting in diagnostic imaging, thereby significantly reducing the workload on physicians. Medical imaging, particularly in the field of radiology, presents unique challenges due to the need for precise localization and classification of pathological findings within high-dimensional and complex visual data. Recent approaches often use pre-trained VLMs on large-scale datasets and then adapt them for downstream tasks such as classification, segmentation, and object detection. However, these approaches are not unified, as they typically focus on individual tasks. For instance, object detection often utilizes the visual encoder component of pre-trained VLMs alongside detection heads like YOLOv3~\cite{wang2022multi, muller2022joint, liu2023m} and R-CNN~\cite{cheng2023prior}, rather than integrating these tasks into a cohesive framework.

To bridge this gap, we propose a novel framework tailored specifically for the medical imaging domain, with a focus on chest X-ray images. Our approach introduces a new task, the \textit{joint localization and classification task}, which aims to simultaneously identify and classify critical findings within medical images. This dual capability is crucial for enhancing diagnostic accuracy and providing comprehensive insights into patient conditions. The VinDr-CXR \cite{nguyen2022VinDr} dataset, comprising 18,000 chest X-ray images annotated by experienced radiologists, serves as the benchmark for our evaluations. This dataset is distinguished by its comprehensive labeling, which includes both local findings (with bounding boxes) and global diagnoses. Our framework capitalizes on this rich annotation to enhance its learning and performance.

In our approach, we integrate multiple visual encoders to capture diverse and detailed representations of the medical images. Specifically, we employ BiomedCLIP~\cite{zhang2023biomedclip} and PubMedCLIP~\cite{eslami-etal-2023-pubmedclip}, two state-of-the-art visual encoders pre-trained on large-scale medical image-text datasets. These encoders extract high-quality visual features that are then projected into the language space of Llama 2, a powerful large language model. The fusion of visual and textual modalities enables our framework to generate detailed and accurate descriptions of medical images, identifying critical findings and their locations.

Our experimental results highlight the superiority of our proposed method over existing approaches. We report significant improvements in both localization accuracy and classification performance, underscoring the potential of our framework to advance the field of medical image analysis. Moreover, the ablation study provides insights into the contributions of various components of our model, offering guidance for future research in this domain.

In summary, this paper presents a comprehensive solution to the challenges of medical image analysis by leveraging the synergistic power of vision and language models. Our framework not only achieves state-of-the-art performance on the VinDr-CXR dataset but also sets a new benchmark for the joint localization and classification of medical images. This paper makes several pivotal contributions, highlighted as follows:

\begin{itemize}
\item We present a new task - \textit{joint localization and classification task} on medical images,
\item We introduce a novel unified vision-language framework designed for the joint task on medical images,
\item We provide fine-tuned models, empirical baselines and conduct an extensive ablation study of our framework, achieving state-of-the-art results on the well-benchmarked VinDr-CXR dataset.
\end{itemize}

All code and models are publicly available online\footnote{https://github.com/leduckhai/LiteGPT}.

\section{Related Work}
\label{sec:related_work}

\subsection{Large Vision-Language Models (VLMs)}

Large Vision-Language Models (VLMs) have significantly advanced capabilities in understanding and generating content that involves both image and textual information. These models are typically pre-trained on datasets containing millions to billions of image-text pairs, and subsequently fine-tuned on specific downstream tasks, enabling these models with the ability to generalize well across various domains. CLIP \cite{radford2021learning} and ALIGN \cite{jia2021scaling} are the pioneering models that have demonstrated remarkable zero-shot performance on computer vision tasks such as image classification and image-text retrieval tasks. Recently, with the advancement of Large Language Models (LLMs) \cite{touvron2023llama, gunasekar2023textbooks, bai2023qwen}, numerous studies \cite{NEURIPS2023_6dcf277e, zhu2023minigpt, chen2023minigptv2, chen2023shikra, liu2024improved, ye2023mplug, ye2024mplug} have leveraged the extensive knowledge embedded in LLMs and the powerful visual understanding capabilities of large vision models like vision transformer ViT. These efforts have resulted in models capable of performing complex reasoning over both text and images, achieving state-of-the-art results in various tasks such as image captioning and visual question answering.

\subsection{Visual Biomedical Question Answering (VBQA)}
Visual Biomedical Question Answering (VBQA) is the task of answering questions related to biomedical images, combining both the complexities of the textual and vision domains. The models in this field are developed with the aim of receiving and understanding medical images such as X-rays, CT scans, MRIs, and other diagnostic visuals. Recently, in the light of multimodal LLMs (MLLMs), several works \cite{li2024llava, moor2023med, thawkar2023xraygpt, he2024pefomed, zhang2023pmc} have emerged, leveraging these MLLMs to enhance VBQA systems. Specifically, a notable work is LLaVA-Med \cite{li2024llava} which is developed based on the LLaVA framework\cite{NEURIPS2023_6dcf277e}. This model employs a two-stage training process to achieve state-of-the-art results across various medical VQA benchmarks, including VQA-RAD \cite{lau2018dataset}, SLAKE \cite{liu2021slake} and PathVQA \cite{he2020pathvqa}. LLaVA-Med is initially pre-trained with 600,000 biomedical image-text pairs and then further trained with instruction-following data generated by GPT-4 to create multi-round questions and answers based on biomedical images and their captions. Another notable model is Med-Flamingo, which is built upon the MLLM Flamingo for biomedical applications and demonstrates highly effective reasoning and in-context learning abilities. These models exemplify the powerful capabilities of MLLMs for the VBQA task.




\section{Method}
\subsection{Model Architecture}

\begin{figure*}[!th]
\centering
\includegraphics[width=\textwidth]{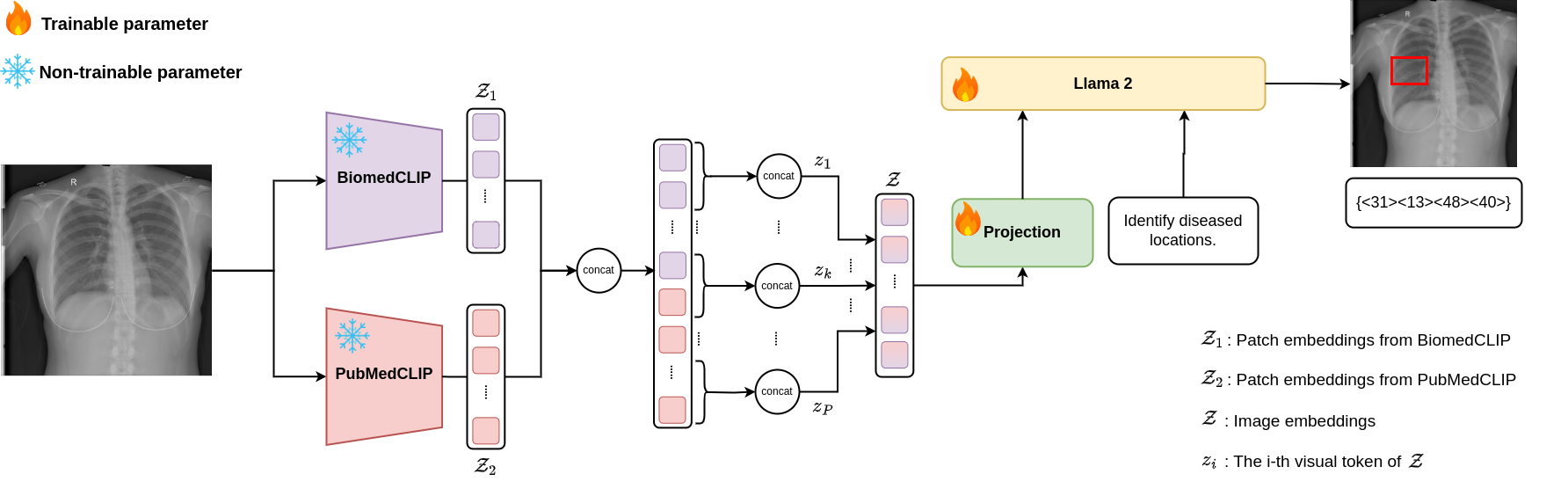}
\caption{\textbf{Overview of our proposed method.} The model employs multiple visual encoders as its visual backbone, specifically incorporating two different encoders, which remain frozen throughout all training phases. We concatenate five adjacent visual tokens from the output of the visual backbone and project them into the language space of Llama 2. The text is embedded through the embedding layer of Llama 2 and directly concatenated with the visual features to generate the answer.
}
\vspace{-8pt}
\label{fig:overview}
\end{figure*}

In this section, we detail our approach, illustrated in Figure \ref{fig:overview}, for identifying critical findings and their corresponding regions, as well as global diseases in chest X-ray images. Inspired by MiniGPT-v2~\cite{chen2023minigptv2},  which efficiently adapts to new vision-language tasks, we have designed a new model adapted for the medical domain, integrating multiple visual encoders.

\subsubsection{Visual Backbone}

In the medical domain, working with unique images such as X-Rays, MRIs, and ultrasounds necessitates the use of advanced visual models. To bridge the capabilities of pre-trained general domain Visual Language Models (VLMs) to the medical domain, we have integrated multiple visual encoders into our visual backbone component. Recognizing that an effective medical model requires high-quality multi-modal data, as well as the limited and costly resources regarding datasets in the medical field, this integration aims to maximize the diversity and comprehensiveness of the information extracted from images.

By leveraging multiple visual encoders, each trained on different large-scale datasets, we can enhance the richness of the image representation. Specifically, we utilize two vision encoders: BiomedCLIP~\cite{zhang2023biomedclip} and PubMedCLIP~\cite{eslami-etal-2023-pubmedclip}. These encoders are trained on extensive datasets containing both medical images and associated text, such as PMC-15M~\cite{zhang2023biomedclip} for BiomedCLIP and ROCO~\cite{roco} for PubMedCLIP.

Initially, to ensure the capabilities of the pre-trained visual encoders are preserved, we set them frozen during the training stage. Each image \(v_i \in \mathbb{R}^{H \times W \times C}\), where \(H\), \(W\), and \(C\) denote the height, width, and channels of the image, respectively, is passed through the pre-trained encoders to extract features, specifically the patch embeddings. Consequently, the representations of the image, $\mathcal{Z}_1$ and $\mathcal{Z}_2$, are obtained from BioMedCLIP and PubMedCLIP as follows:
\begin{equation}
    \mathcal{Z}_1 = \text{VE}_{\text{BiomedCLIP}}(v_i) \in \mathbb{R}^{P_1 \times\ 768},
\end{equation}
\begin{equation}
    \mathcal{Z}_2 = \text{VE}_{\text{PubMedCLIP}}(v_i) \in \mathbb{R}^{P_2 \times\ 768},
\end{equation}


\noindent where \(P_1\) and \(P_2\) are the number of patches of BiomedCLIP and PubMedCLIP, respectively. We then concatenate $\mathcal{Z}_1$ and $\mathcal{Z}_2$ to obtain the unified presentation of the image as follows:
\begin{equation}
    \mathcal{Z} = \text{Concatenate}(\mathcal{Z}_1, \mathcal{Z}_2) \in \mathbb{R}^{P \times\ 768},~P = P_1 + P_2. \label{eq3}
\end{equation}

\subsubsection{Visual Projection}
This component aims to project the visual features \(Z\) obtained from Eq. \eqref{eq5} into the language model space. Specifically, the projection consists of two linear layers, with a GELU activation function~\cite{hendrycks2016gaussian} between them. Notably, for the final linear layer, we leverage the pre-trained weights of the linear layer from MiniGPT-v2, which connects the vision backbone to the language model. To optimize efficiency during training and inference, inspired by MiniGPT-v2, we concatenate five consecutive visual tokens before inputting them into the projection. This method increases the dimensionality of each visual token by a factor of five while reducing the number of tokens by the same factor, thereby minimizing the input size to the language model.

Following Eq. \eqref{eq3}, we have the visual representations $\mathcal{Z} = [z_1, z_2, ..., z_P] \in \mathbb{R}^{P \times\ 768}$. After concatenating every five adjacent visual tokens into a single token, the resulting visual representations $\mathcal{Q} = [q_1, q_2, ..., q_M] \in \mathbb{R}^{M \times\ 3840}$, where $M = \frac{P}{5}$ and each $q_i (1 \leq i \leq M)$ is defined as follows:
\begin{equation}
    q_i = \text{Concatenate}(z_{5i-4}, z_{5i-3}, z_{5i-2}, z_{5i-1}, z_{5i}).
\end{equation}

The visual projection layer \(f\) is applied to map $\mathcal{Q} \in \mathbb{R}^{M \times\ 3840}$ to $\mathcal{V} \in \mathbb{R}^{M \times\ D}$ where $D$ represents the hidden size of the language model, as follows:
\begin{equation}
    \mathcal{V} = f(\mathcal{Q}) \in \mathbb{R}^{M \times\ D}. \label{eq5}
\end{equation}

\begin{figure*}[h]
  \centering
   \includegraphics[width=0.9\textwidth]{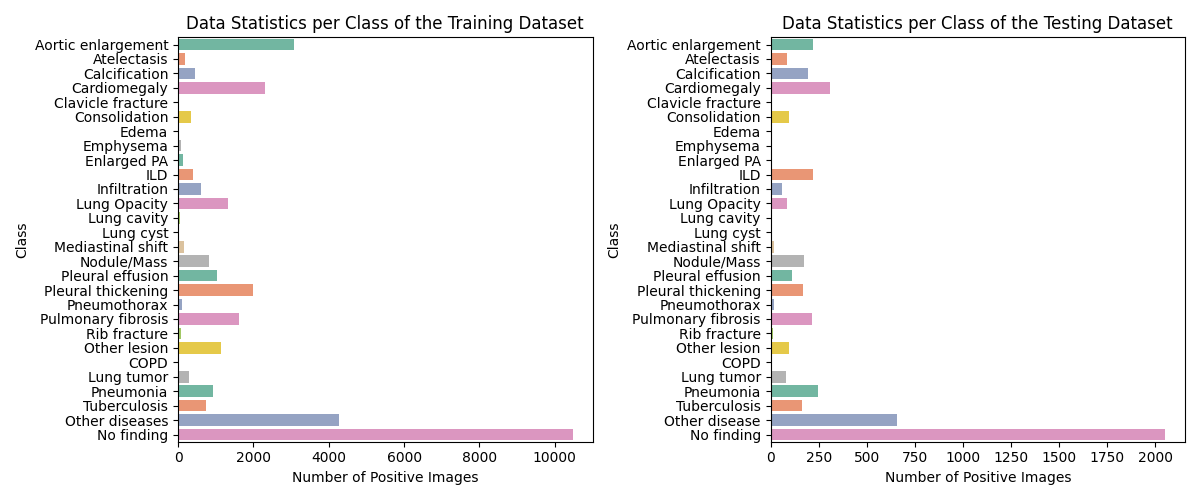}
   \caption{\textbf{The distribution for a total of 28 findings and diagnoses.} The numbers of positive labels were reported based on the assessments of the participating radiologists. 
   }
   \label{fig:distribution_dataset}
\end{figure*}

\subsubsection{Large Language Model}

In the context of vision-language models, the language is pivotal, while the vision serves merely as additional information to guide the language model in generating answers. In this paper, one of the main tasks is the localization of chest X-rays. Therefore, we need a language model adept at identifying the spatial regions of anomalies within the images. To achieve this, we selected Llama 2-Chat (7B)~\cite{llama2}, which is pre-trained on a variety of vision-language tasks via MiniGPT-v2. Specifically, we initialize the pre-trained weights of the Llama 2-Chat component from MiniGPT-v2. We refer to this language model as the foundation for transferring knowledge into the medical domain we are working on. 

The text is processed through a word embedding layer to obtain sequence tokens $\mathcal{T} \in \mathbb{R}^{N \times\ D}$, where $N$ represents the sequence length of text tokens. Subsequently, the visual tokens from Eq. \eqref{eq5} are concatenated with the text tokens $\mathcal{T}$ and fed into Llama 2-Chat to generate answers.

\subsection{Training Stages}

The training process consists of two stages: the first is grounded critical findings learning, and the second is diagnosis learning.

\textbf{Grounded critical findings learning stage.} In this stage, our objective is to adapt the language model to the medical domain, enabling it to comprehend anomalies in X-ray scans. We select samples that contain identified anomalies along with their specific locations in the image. The language model is then directly instructed to perform grounded critical findings detection. 

\textbf{Diagnose learning stage.} After the first stage, the model acquires sufficient knowledge to identify regions in the image that contain anomalies. In this stage, the model utilizes this acquired knowledge to diagnose the diseases present in the image. The emphasis is on leveraging the detailed understanding of the localized findings to provide accurate and comprehensive diagnoses.

\subsection{Input Template}

The general input format consists of three parts: the first part is the image input, the second part is the identifier token, and the third part is the instruction input. Details of each part are provided below.

\textbf{Image input}. The images are processed through the visual backbone and projection to obtain a fixed-length sequence for input into the language model. To differentiate between tokens derived from images and those from text, two special tokens, \verb|<image>| and \verb|</image>|, are used to concatenate the beginning and end of the image feature sequence, respectively. This indicates where the image content starts and ends.

\textbf{Identifier token}. In the first training stage, we define our localization task as a grounded captioning task. Specifically, we employ the \verb|[identify]| token, which is pre-trained for visual grounding within the MiniGPT-v2 framework, and adapt it to our medical domain localization task. The rationale for choosing this token is detailed in section \ref{ablation_study}. For the second training stage, we define our task as a visual question answering (VQA) task. Therefore, we use the \verb|[vqa]| token and construct instructions tailored to the VQA task. The detailed instruction is presented in section \ref{expersetup}. Overall, separating the two tasks helps the model easily differentiate between them and improves learning efficiency.

\textbf{Instruction input}. The instruction input provides specific commands or queries that guide the language model's response generation. These instructions can vary based on the desired task, such as generating captions, answering questions, or identifying specific regions in the image. The input is formatted as natural language text, enabling the model to understand and execute the given instructions effectively. The specific instructions we use are detailed in Section \ref{expersetup}.

\section{Experiments}
\subsection{Dataset}
In this study, we utilize the VinDr-CXR dataset~\cite{nguyen2022VinDr}, gathered from Hospital 108 and Hanoi Medical University Hospital, two of the leading hospitals in Vietnam. The released dataset comprises 18,000 chest X-ray images in DICOM format, all captured in the Posterior-Anterior (PA) view. A team of 17 experienced radiologists manually annotated the images, identifying 22 critical findings (local labels), with each finding localized using bounding boxes, and 6 diagnoses (global labels). The dataset is divided into a training set and a testing set. The training set includes 15,000 images, each independently labeled by 3 radiologists, while the testing set contains 3,000 images, meticulously labeled by a consensus of 5 radiologists. The detailed distribution of the number of positive images in both the training and testing datasets is shown in Figure \ref{fig:distribution_dataset}. 

The VinDr-CXR is one of the few datasets that include labels from a list of findings specifying their locations on the radiographs and is the largest dataset in terms of the number of findings.


\subsection{Data Preprocessing}
The raw image data is taken and processed by normalizing color values to 8 bits from the variable 10-16-bit color space that the images were originally stored in. The normalization was done either through the stored windowing parameters or if they were missing, the full-color space of the original image. 
Images with conflicting labels, such as one radiologist labeling an image as \textit{no finding} while another provides a full label, are removed from the dataset.

For each bounding box, we normalize the coordinates, defined as $\left[\frac{x_{min}}{w}, \frac{y_{min}}{h}, \frac{x_{max}}{w}, \frac{y_{max}}{h}\right]$, where \(x\) and \(y\) denote the coordinates, and \(h\) and \(w\) denote the height and width of the image, respectively. Each coordinate is multiplied by 100 to obtain integer values within the range [0,100]. These normalized coordinates are then transformed into the string format ``$\{<x_{min}><y_{min}><x_{max}><y_{max}>\}$", where \(x_{min}\) and \(y_{min}\) represent the \(x\) and \(y\) coordinates of the top-left corner, and \(x_{max}\) and \(y_{max}\) represent the \(x\) and \(y\) coordinates of the bottom-right corner of the bounding box. Moreover, multiple nearly identical bounding boxes often appear in an image for the same disease, implying that one disease can have multiple overlapping regions. We define these overlapping regions as having a pairwise Intersection over Union (IoU) higher than 0.5. To address this, we select one representative bounding box and remove the remaining redundant regions.

\subsection{Evaluation Metrics}

We report the classification performance using precision, recall, and F1-score. The evaluation is conducted by comparing the model’s response to the ground truth. Specifically, a class present in both the ground truth and the model’s response is considered correct for the classification report.

Regarding the localization task, performance is evaluated based on accuracy. Specifically, for each critical finding (local labels) present in the ground truth that matches the model's response, we consider the predicted bounding box accurate if its IoU with the ground truth exceeds the specified threshold.

For text validity evaluation, we employ common natural language processing metrics: ROUGE, BLEU, METEOR, and CIDEr scores. The generated text undergoes modification by excluding the localization details, as the bounding box coordinates may not reflect the model’s summarization capabilities effectively.

\begin{table*}[!t]
\centering
\resizebox{0.8\textwidth}{!}{%
\begin{tabular}{c|cc|cc|cc}
\hline
\multicolumn{1}{c|}{\multirow{2}{*}{\textbf{Method}}} & \multicolumn{2}{c|}{\textbf{Precision (\(\%\))}}            & \multicolumn{2}{c|}{\textbf{Recall (\(\%\))}}               & \multicolumn{2}{c}{\textbf{F1-score (\(\%\))}}             \\ \cline{2-7} 
\multicolumn{1}{c|}{}                        & \multicolumn{1}{c|}{Finding} & No Finding & \multicolumn{1}{c|}{Finding} & No Finding & \multicolumn{1}{c|}{Finding} & No Finding \\ \hline
MiniGPT-v2~\cite{chen2023minigptv2}                          & \multicolumn{1}{c|}{73}       & \textbf{83}          & \multicolumn{1}{c|}{\textbf{59}}       & 90          & \multicolumn{1}{c|}{66}       & \underline{86}          \\ 
BiomedCLIP - Llama 2                          & \multicolumn{1}{c|}{\textbf{88}}       & \underline{82}          & \multicolumn{1}{c|}{\underline{54}}       & \textbf{97}          & \multicolumn{1}{c|}{\underline{67}}       & \textbf{89}          \\ 
PubMedCLIP - Llama 2                          & \multicolumn{1}{c|}{73}       & 80          & \multicolumn{1}{c|}{49}       & 92          & \multicolumn{1}{c|}{59}       & 85          \\ 
BiomedCLIP + PubMedCLIP - Llama 2             & \multicolumn{1}{c|}{\underline{84}}       & \textbf{83}          & \multicolumn{1}{c|}{\textbf{59}}       & \underline{95}          & \multicolumn{1}{c|}{\textbf{70}}       & \textbf{89}          \\ \hline           
\end{tabular}}
\caption{\textbf{Results on Classification Task}. We evaluated the model's performance using the prompt of \textit{[vqa] Given the provided chest X-ray image, which of the following diagnoses are present (select all that apply): COPD, Lung Tumor, Pneumonia, Tuberculosis, Other Disease, or No Finding?} \textbf{Bold} indicates the best result, and \underline{underline} indicates the second-best result.}
\label{tab1}
\end{table*}

\begin{table}[!t]
\adjustbox{max width=\columnwidth}
\centering
\resizebox{\columnwidth}{!}{%
\begin{tabular}{cc}
\hline
    \textbf{Method}                            & \textbf{F1-score (\(\%\))} \\ \hline
    DeViDe~\cite{luo2024devide} (*)                            & 41.28    \\
    KAD~\cite{zhang2023knowledge} (*)                              & 40.22    \\
    DWARF~\cite{luo2024dwarf} (*)                           & 47.05    \\
    MiniGPT-v2~\cite{chen2023minigptv2} (**)                           & 53    \\ \hline
    BiomedCLIP - Llama 2              & \underline{59}        \\
    PubMedCLIP - Llama 2              & 55        \\
    BiomedCLIP + PubMedCLIP - Llama 2 & \textbf{60}        \\ \hline
\end{tabular}}
\caption{\textbf{Results on Classification Task}. (*) indicates a pre-trained vision-language model without fine-tuning, while (**) indicates that the model has been fine-tuned on the VinDr-CXR dataset. \textbf{Bold} indicates the best result, and \underline{underline} indicates the second-best result.
}\label{tab2}
\end{table}







\subsection{Experimental Settings} \label{expersetup}

The instructional input used during the \textbf{first training stage} is: ``\textit{Please describe the critical findings along with their localized bounding boxes in the radiological image of a chest as much detail as possible. If there are no findings, state that the chest radiograph shows no findings.}". For the \textbf{second training stage}, the instructional input is: ``\textit{Given the provided chest X-ray image, which of the following diagnoses are present (select all that apply): COPD, Lung Tumor, Pneumonia, Tuberculosis, Other Disease, or No Finding?}"


In our experiments, we use the AdamW optimizer~\cite{loshchilov2018decoupled} with a learning rate scheduler that includes a linear warmup followed by a cosine decay. The initial learning rate is set to 1e-5, with a minimum learning rate of 1e-5 and a warmup learning rate of 1e-5. The warmup phase consists of 1000 steps. We apply a weight decay of 0.05. The training process utilizes 20 epochs, and each epoch consists of 549 iterations. The training is conducted using 2 workers. The input image resolution was 224 $\times$ 224. The model training was conducted on a computing system equipped with four A100 GPUs and took approximately three hours. We trained the model using LoRA~\cite{hu2022lora}, a parameter-efficient fine-tuning technique. The model was loaded in 8-bit precision, and the $\mathcal{W}_q$ and $\mathcal{W}_v$ parameters were initialized from MiniGPT-v2. The rank, \(r\), was set to 64. We continued fine-tuning these parameters via lower-rank adaptation.

\subsection{Experimental Results}
\subsubsection{Classification Performance}

In this section, we first present the performance of binary classification assessed using an anomaly detection framework to evaluate the model's ability to detect anomalous samples. We compare the results with the vision-language baselines that we trained on the VinDr-CXR dataset, including BiomedCLIP + PubMedCLIP - Llama 2, BiomedCLIP - Llama 2, PubMedCLIP - Llama 2, and MiniGPT-v2, to determine which model performs better in the medical domain. Precision, Recall, and F-1 scores for both the \textit{finding} and \textit{no finding} classes are reported in Table \ref{tab1}. As shown in Table \ref{tab1}, the recall for the \textit{no finding} class exceeds 90\(\%\), which is attributed to the predominance of \textit{no finding} samples in both the training and testing datasets. In contrast, the recall for the \textit{finding} class is only about 59\(\%\), due to the limited and imbalanced dataset, whereas the \textit{no finding} class accounts for about 67\(\%\) of the total dataset. This imbalance hinders the model's ability to learn and classify effectively, introducing bias in classification. Overall, the BiomedCLIP + PubMedCLIP - Llama 2 model slightly outperforms other baselines. Notably, it demonstrates strong performance in detecting anomalies in each X-ray scan, as evidenced by a recall rate of 59\(\%\) and a high precision rate of 84\(\%\).

\begin{table*}[!t]
\centering
\resizebox{0.7\textwidth}{!}{%
\begin{tabular}{cccc}
\hline
\textbf{Method}      & \textbf{Accuracy@0.3} (\(\%\))                      & \textbf{Accuracy@0.4} (\(\%\)) & \textbf{Accuracy@0.5} (\(\%\)) \\ \hline
MiniGPT-v2~\cite{chen2023minigptv2}       & 5.9       & 5.01            & 3.38            \\
BiomedCLIP - Llama 2          & \underline{10.72}   & \underline{9.01}            & \underline{6.75}            \\
PubMedCLIP - Llama 2          &  6.23  & 4.38            & 2.86            \\
BiomedCLIP + PubMedCLIP - Llama 2   & \textbf{12.54}  & \textbf{10.5}            & \textbf{8.12}          \\ \hline 
\end{tabular}}
\caption{\textbf{Results on Localization Task}. We evaluated the model's performance using the prompt of \textit{[identify] Please describe the critical findings along with their localized bounding boxes in the radiological image of a chest in as much detail as possible.} \textbf{Bold} indicates the best result, and \underline{underline} indicates the second-best result.}
\label{tab3}
\end{table*}

\begin{table*}[!t]
\adjustbox{max width=\textwidth}
\centering
\resizebox{\textwidth}{!}{%
\begin{tabular}{ccccccccccc}
\hline
\textbf{Method}                   & \multicolumn{1}{c}{\textbf{ROUGE-1}} (\(\%\)) & \multicolumn{1}{c}{\textbf{ROUGE-2}} (\(\%\))& \multicolumn{1}{c}{\textbf{ROUGE-L}} (\(\%\))& \multicolumn{1}{c}{\textbf{ROUGE-LSUM}} (\(\%\))& \multicolumn{1}{c}{\textbf{BLEU-1}} (\(\%\))& \multicolumn{1}{c}{\textbf{BLEU-2}} (\(\%\))& \multicolumn{1}{c}{\textbf{BLEU-3}} (\(\%\))& \multicolumn{1}{c}{\textbf{BLEU-4}} (\(\%\))& \multicolumn{1}{c}{\textbf{METEOR}} (\(\%\))& \multicolumn{1}{c}{\textbf{CIDEr}} \\ \hline
MiniGPT-v2                        & 42.85                                & 35.37                                & 42.28                                & 42.27                                   & 21.84                               & 20.6                                & 19.4                                & 18.17                               & 61.54                               & 0.29                               \\
BiomedCLIP - Llama 2              & 49.51                                & 38.77                                & 48.5                                 & 48.46                                   & \textbf{32.76}                      & \textbf{30.56}                      & \textbf{28.62}                      & \textbf{26.71}                      & \textbf{68.73}                      & \underline{ 0.46}                         \\
PubMedCLIP - Llama 2              & 22.3                                 & 14.31                                & 21.58                                & 21.6                                    & 14.16                               & 12.95                               & 11.95                               & 10.98                               & 34                                  & 0.1                                \\
BiomedCLIP + PubMedCLIP - Llama 2 & \textbf{50.1}                        & \textbf{40.58}                       & \textbf{48.75}                       & \textbf{48.74}                          & \underline{ 28.96}                         & \underline{ 27.02}                         & \underline{ 25.37}                         & \underline{ 23.74}                         & \underline{ 61.92}                         & \textbf{0.56} \\ \hline                     
\end{tabular}}
\caption{\textbf{Results on text validity for the localization task}. We compute the common metrics of natural language processing based on text generated from LLM. \textbf{Bold} indicates the best result, and \underline{underline} indicates the second-best result.}
\label{tab4}
\end{table*}

\begin{table*}[!t]
\adjustbox{max width=\textwidth}
\centering
\resizebox{\textwidth}{!}{%
\begin{tabular}{ccccccccccc}
\hline
\textbf{Method}                   & \textbf{ROUGE-1} (\(\%\))& \textbf{ROUGE-2} (\(\%\))& \textbf{ROUGE-L} (\(\%\))& \textbf{ROUGE-LSUM}  (\(\%\))& \textbf{BLEU-1} (\(\%\))& \textbf{BLEU-2} (\(\%\))& \textbf{BLEU-3} (\(\%\))& \textbf{BLEU-4} (\(\%\))& \textbf{METEOR} (\(\%\))& \textbf{CIDEr}      \\ \hline
MiniGPT-v2                        & 94.48            & 93.35            & 94.47            & 94.47                & 94.22           & 92.32           & 91.15           & 90.05           & 93.91           & 7.26                \\
BiomedCLIP - Llama 2              & \textbf{95.18}   & \textbf{94.14}   & \textbf{95.18}   & \textbf{95.18}       & \textbf{94.43}  & \textbf{92.69}  & \textbf{91.61}  & \textbf{90.59}  & \textbf{94.8}   & \textbf{7.57}       \\
PubMedCLIP - Llama 2              & 93.31            & 91.80            & 93.31            & 93.31                & 88.59           & 86.33           & 84.82           & 83.34           & 92.58           & 7.06                \\
BiomedCLIP + PubMedCLIP - Llama 2 & \underline{ 95.12}      & \underline{ 94.08}      & \underline{ 95.12}      & \underline{ \textbf{95.12}} & \textbf{94.45}  & \textbf{92.70}  & \underline{ 91.60}     & \underline{ 90.56}     & \underline{ 94.74}     & \underline{ \textbf{7.52}} \\ \hline
\end{tabular}}
\caption{\textbf{Results on text validity for the classification task}. We compute the common metrics of natural language processing based on text generated from LLM. \textbf{Bold} indicates the best result, and \underline{underline} indicates the second-best result.}
\label{tab5}
\end{table*}

\begin{table}[hbt!]
\adjustbox{max width=\columnwidth}
\centering
\resizebox{\columnwidth}{!}{%
\begin{tabular}{c|ccc}
\hline
\textbf{Identifier} & \multicolumn{1}{c}{\textbf{Accuracy@0.3}} & \multicolumn{1}{c}{\textbf{Accuracy@0.4}} (\(\%\))& \multicolumn{1}{c}{\textbf{Accuracy@0.5}} (\(\%\))\\ \hline
{[}grounding{]}    &            11.98                               &                     10.31                      &          8.04                                 \\
{[}identify{]}    &                  \textbf{12.54}                         &                                \textbf{10.5}           &     \textbf{8.12}           \\ \hline                         
\end{tabular}}
\caption{\textbf{Results on Ablation Study}. The impact of different identifier tokens.}
\label{ablation_study}
\end{table}


Secondly, we establish baselines for comparison, including pre-trained vision-language models without fine-tuning, such as DeViDe, and KAD, as well as DWARF, and MiniGPT-v2 fine-tuned on the VinDr-CXR dataset. The results, shown in Table \ref{tab2}, highlight the differences in F1-scores across various models on a classification task. The pre-trained models DeViDe and KAD achieve F1-scores of 41.28\(\%\) and 40.22\(\%\), respectively. In contrast, the fine-tuned models yield more impressive results, with MiniGPT-v2 achieving 53\(\%\), BiomedCLIP - Llama 2 at 59\(\%\), and PubMedCLIP - Llama 2 at 55\(\%\). Notably, the combination of BiomedCLIP and PubMedCLIP with Llama 2 outperforms all others, achieving an F1-score of 60\(\%\), an improvement from 5\(\%\) to 19.78\(\%\). This demonstrates the effectiveness of fine-tuning and model integration.

\subsubsection{Localization Performance}

In this study, we evaluated the performance of our model in generating bounding boxes for critical findings. The model is required to accurately classify 22 different anomalies on the X-ray scan and then determine their locations. Only correctly classified findings are included in the evaluation metrics. However, there is a notable absence of pure vision-language models addressing localization tasks in the medical domain. Typically, these tasks use a pre-trained vision encoder as the backbone along with a detection head module. Our review of recent literature indicates that localization tasks in VinDr-CXR follow a similar approach. These methods are primarily evaluated using mean Average Precision (mAP), as each bounding box includes a confidence value, which facilitates the computation of mAP. In contrast, bounding boxes generated by vision-language models are in pure text format and lack confidence values, making mAP comparisons challenging. Therefore, in this paper, we establish our model as the baseline for the first vision-language model addressing localization tasks in the medical domain.

The results of our experiments are summarized in Table \ref{tab3}. The metric Accuracy@threshold indicates that a predicted bounding box is deemed correct if its IoU with the ground truth bounding box exceeds the specified threshold. In this experiment, we used thresholds of 0.3, 0.4, and 0.5. We compared our established baselines to a robust multi-modal large language model, MiniGPT-v2, which serves as a unified interface for various vision-language multi-task learning, including localization tasks. We chose MiniGPT-v2 as a baseline for our experiment based on the findings of Jun Chen et al.~\cite{chen2023minigptv2}, which indicated that MiniGPT-v2 efficiently adapts to new vision-language tasks. As shown in the table, the model that integrates multiple visual encoders outperforms those with a single encoder and MiniGPT-v2 fine-tuned on the VinDr-CXR dataset, achieving an Accuracy@0.3 of 12.54\(\%\), an Accuracy@0.4 of 10.5\(\%\), and an Accuracy@0.5 of 8.12\(\%\). Compared to MiniGPT-v2, these results show significant improvements of 6.64\(\%\) for Accuracy@0.3, 5.49\(\%\) for Accuracy@0.4, and 4.74\(\%\) for Accuracy@0.5. Notably, the integration of multiple visual encoders enhances the richness of image information, allowing the model to identify locations more effectively, resulting in higher performance than models with a single visual encoder.

\subsubsection{Text Validity Performance}

In addition to the main metrics of our task, we also evaluate the performance of the text generated by the model. Tables \ref{tab4} and \ref{tab5} compare four methods across various text evaluation metrics for classification and localization tasks. Overall, BiomedCLIP + PubMedCLIP - Llama 2 achieves the highest scores in ROUGE and CIDEr metrics, indicating superior performance in text validity for both classification and localization tasks. BiomedCLIP - Llama 2 also performs well, leading in BLEU and METEOR scores. The combination of BiomedCLIP and PubMedCLIP outperforms the other methods, showcasing the benefit of integrating multiple encoders.


\subsection{Ablation study}
\label{ablation_study}

In this ablation study, we explore the effect of different identifier tokens on localization task performance. Specifically, we selected the [grounding] and [identify] tokens, which perform well in the MiniGPT-v2 model. Both tokens were trained under the same conditions, including the VinDr-CXR dataset, instructional input, and experimental settings. The results, detailed in Table \ref{ablation_study}, show that the identifier token \verb|[identify]| outperforms \verb|[grounding]|, achieving better accuracy scores at thresholds of 0.3, 0.4, and 0.5. This can be attributed to the [identify] token being trained on the referring expression generation (REG) task with various datasets (RefCOCO~\cite{kazemzadeh-etal-2014-referitgame}, RefCOCO+~\cite{yu2016modeling}, and RefCOCOg~\cite{mao2016generation}) over three stages of training, enabling it to effectively learn spatial relationships in images. In contrast, the [grounding] token was trained on the grounded caption task using only the GRIT-20M dataset from KOSMOS-v2~\cite{peng2023kosmos} in a single stage, which did not provide sufficient knowledge for learning spatial relationships.

\section{Qualitative results}

\begin{table*}[hbt!]
\centering  
\small
\begin{tabular}{cp{2cm}p{10cm}  }
\toprule
\multicolumn{3}{l}{\bf Example 1.}  \\
\midrule
\multirow{3}{*}{ \includegraphics[width=2.2cm, height=2.2cm]{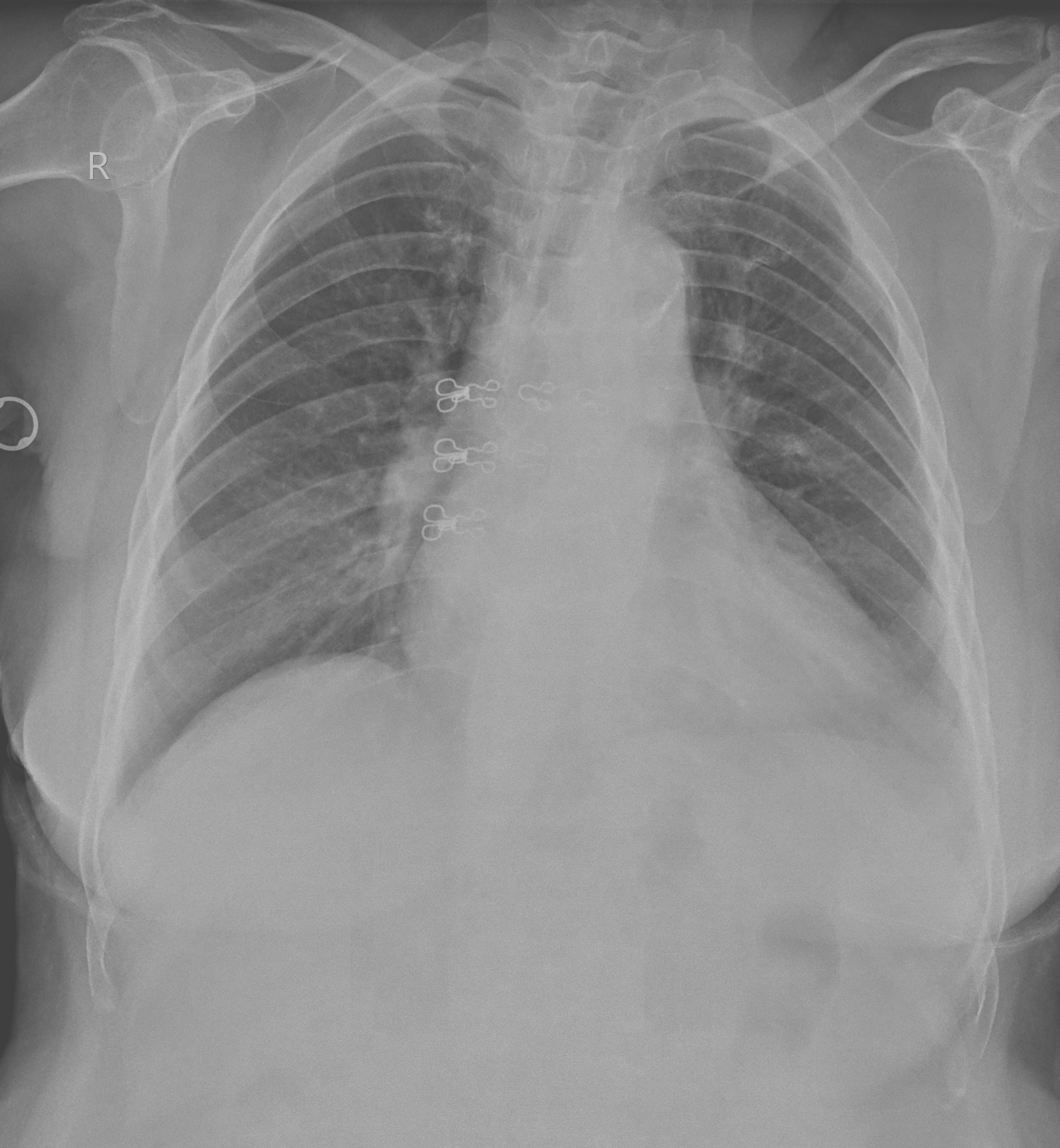}  }
& Predict & Local diseases of this chest radiograph are $<$p$>$Aortic enlargement$<$/p$>$ $\{<56><17><67><28>\}$,\textcolor{green!70!black}{$<$p$>$Cardiomegaly$<$/p$>$} $\{<38><48><85><65>\}$.
\\
& Ground truth & Local diseases of this chest radiograph are $<$p$>$Calcification$<$/p$>$ $\{<60><21><66><29>\}$,\textcolor{green!70!black}{$<$p$>$Cardiomegaly$<$/p$>$} $\{<35><50><86><67>\}$.

\\
\toprule
\multicolumn{3}{l}{\bf Example 2.}  \\
\midrule
\multirow{3}{*}{ \includegraphics[width=2.2cm, height=2.2cm]{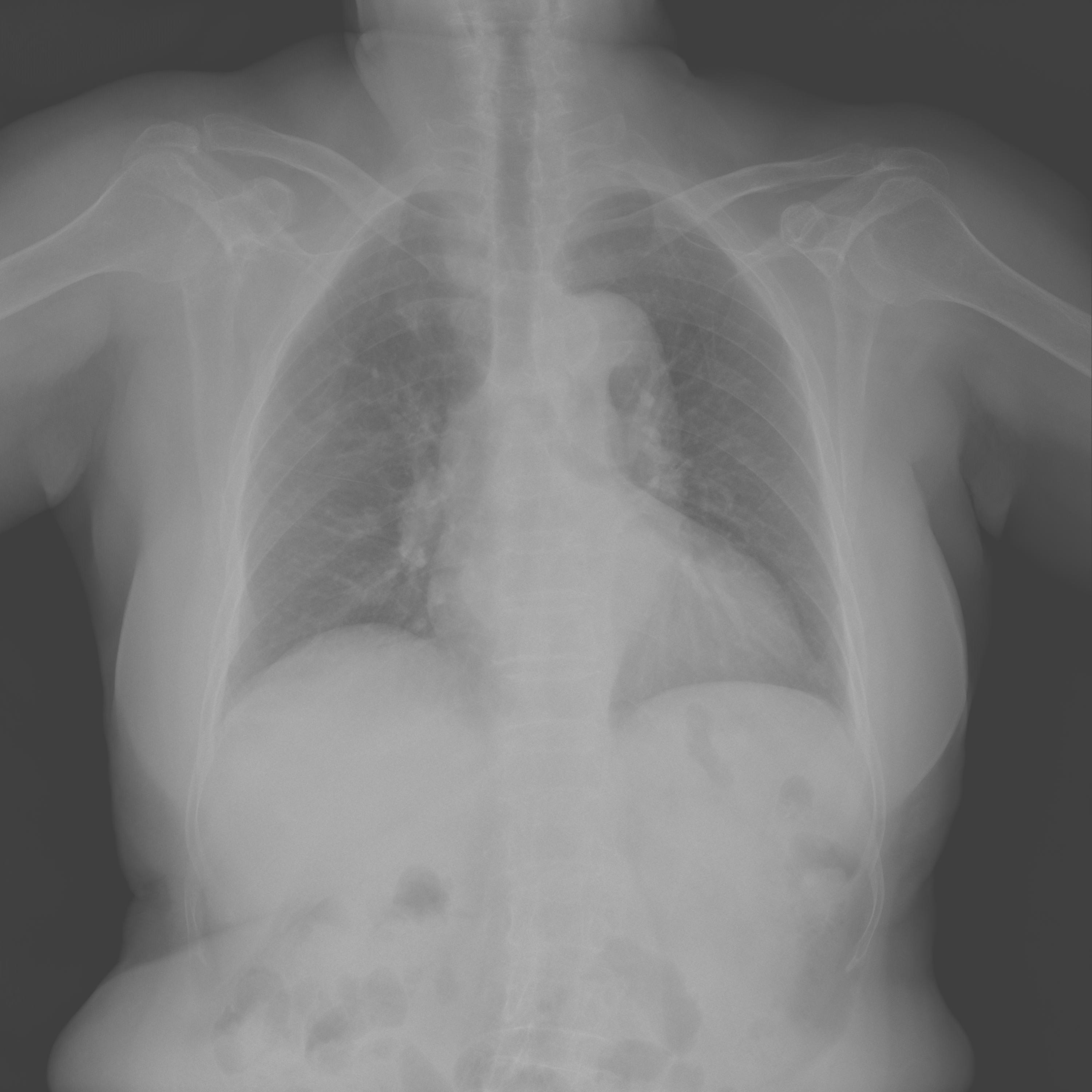}  }
&  Predict &  Local diseases of this chest radiograph are \textcolor{green!70!black}{$<$p$>$Aortic enlargement$<$/p$>$} $\{<48><25><60><36>\}$,\textcolor{green!70!black}{$<$p$>$Cardiomegaly$<$/p$>$} $\{<42><51><74><62>\}$.
\\
& Ground truth &  Local diseases of this chest radiograph are \textcolor{green!70!black}{$<$p$>$Cardiomegaly$<$/p$>$} $\{<38><48><75><65>\}$,\textcolor{green!70!black}{$<$p$>$Aortic enlargement$<$/p$>$} $\{<39><27><64><48>\}$.  \\
\bottomrule
\end{tabular}
\caption{\textbf{Examples of Localization Task on Radiology Images}: The text in \textcolor{green!70!black}{Green} indicates anomalies classified correctly.}
\label{tab7}
\end{table*}

\begin{figure*}[hbt!]
  \centering
   \includegraphics[width=0.6\textwidth]{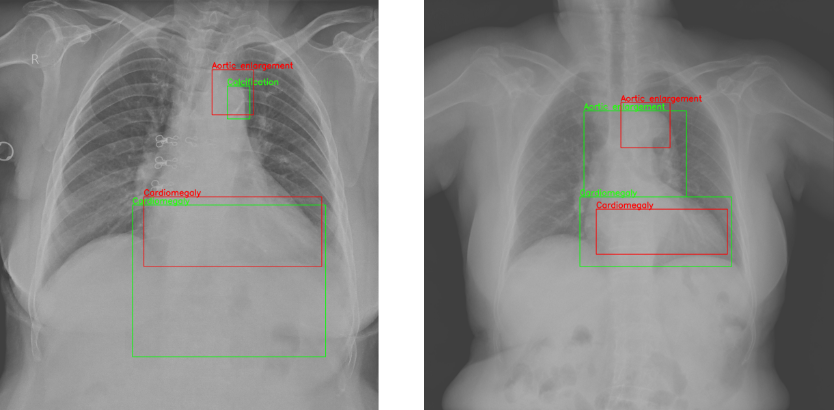}
   \caption{\textbf{The visualization of bounding boxes is shown in Table \ref{tab7}}. The red rectangles indicate the model's predicted boxes, while the green rectangles indicate the ground truth boxes. In the image on the \textbf{left} side, representing example 1, the red box at the top represents Aortic enlargement, the green box at the top represents Calcification, and the boxes at the bottom represent Cardiomegaly. In the image on the \textbf{right} side, representing example 2, the red and green boxes at the top represent Aortic enlargement, and the boxes at the bottom represent Cardiomegaly.}
   \label{fig:examples_bboxes}
\end{figure*}

In this section, we present quantitative results to demonstrate the effectiveness of our method. The generated results are illustrated in Table \ref{tab7}, and the visualization of the bounding boxes is shown in Figure \ref{fig:examples_bboxes}. 

As illustrated in Example 1, we could easily observe that Cardiomegaly is detected correctly, and the predicted region closely matches the ground truth. Additionally, while the model incorrectly classified Calcification and Aortic Enlargement, the predicted bounding box aligns with the ground truth. As shown in Example 2, the results are more accurate, with both Aortic Enlargement and Cardiomegaly correctly classified, and the regions indicated by these detections are acceptable. Overall, diagnosing and detecting anomolies in medical images is challenging not only for radiologists but also for advanced artificial intelligence systems due to variability in diagnoses resulting from different perspectives of radiologists. These models should serve as assistants, providing additional information and recommendations to support radiologists in making decisions.

\section{Conclusion}
In this work, we present the first unified framework for a large vision-language model to address the new task - joint localization and classification task in medical imaging. Our approach leverages multiple visual encoders rather than a single visual encoder, resulting in superior performance in disease localization and image classification, as well as across nearly all text validity metrics. This is shown using the well-benchmarked VinDr-CXR dataset, where our model achieves state-of-the-art results on the classification task. We also find that the identifier token [identify] exhibits superior performance compared to the [grounding] token in the localization task, due to the [grounding] token's insufficient knowledge base for learning spatial relationships. Furthermore, we establish baselines for the localization task within this dataset, providing valuable benchmarks for future research on unified vision-language models in the context of medical imaging.

\section{Acknowledgement}
This work was partially done as an undergraduate thesis of Ryan Zhang under the supervision of Khai Le-Duc at University of Toronto.

{\small
\bibliographystyle{ieee_fullname}
\bibliography{egbib}
}

\clearpage
\newpage
\onecolumn

\tableofcontents
\newpage

\section*{Appendix}
\appendix
\section{Details of Data Preprocessing}
\label{sec:details_data_preprocessing}

\subsection{DICOM to RGB Conversion}
Converting from DICOM (Digital Imaging and Communications in Medicine) to RGB (Red, Green, Blue) can reduce disk space due to several reasons related to data compression and file format efficiency:

\begin{itemize}
    \item Compression Algorithms: Many image formats that use RGB (such as JPEG or PNG) support various compression algorithms. JPEG, for instance, uses lossy compression, which can significantly reduce the file size by discarding some of the less important image information \cite{wiseman2015still}. Even lossless compression algorithms used in formats like PNG can be more efficient than the compression used in DICOM files \cite{lin1997robust}.
    \item Metadata Overhead: DICOM files often contain extensive metadata about the image, patient, and imaging parameters. This metadata, while crucial in medical contexts, adds to the file size \cite{caffery2018transforming, blackledge2014stegacryption, clunie2021dicom}. When converting to a standard RGB image format, much of this metadata is discarded or simplified, resulting in smaller file sizes \cite{fidler2006lossy, varma2012managing}.
    \item Pixel Data Encoding: DICOM files typically use higher bit depths and may store additional information per pixel (e.g., 16-bit grayscale images), whereas standard RGB images usually use 8 bits per color channel (24 bits total) \cite{chen2012study, pianykh2012medical, liu2007medical, manikandan2021dual}. Reducing the bit depth can decrease the file size.
    \item Specialized Compression Techniques: DICOM may use specialized medical image compression methods that are not as efficient in terms of disk space as general-purpose image formats \cite{cazanas2022digital, european2011usability, fritsch2011lossy}.
\end{itemize}

A DICOM file with a 16-bit grayscale image might be several megabytes in size due to the high bit depth and extensive metadata \cite{goel2008mathematical, lebre2021efficient, aryanto2021performance, elhadad2021blind, trieu2023use, lang2023dicom, stewart1999integration, lamy2015design}. Converting this image to an 8-bit per channel RGB JPEG file can reduce the size drastically, especially when using lossy compression. However, it is crucial to consider that converting medical images from DICOM to RGB may result in the loss of important clinical information, reduced image quality, and the inability to use specialized medical image analysis tools \cite{herrmann2018implementing, lalitha2010lossless, pianykh2012medical, olges2000integrating}. 

Below is the Python code for converting DICOM to JPG format.

\begin{python}
image_path = os.path.join(root, filename)
image_dicom = pydicom.read_file(image_path)
image = image_dicom.pixel_array.astype(np.float64)
    
if image_dicom[0x00280004].value == "MONOCHROME1":
    flip = -1
else:
    flip = 1
try:
    normalized_image = ((image - float(image_dicom[0x00281050].value)) / float(image_dicom[0x00281051].value))
    image = ((np.clip(normalized_image, -1, 1) * flip * 127.5 + 127.5)).astype(np.uint8)
except KeyError:
    normalized_image = ((image - np.mean(image)) / (np.max(image) - np.min(image)))
    image = (np.clip(normalized_image, -1, 1) * flip * 127.5 + 127.5).astype(np.uint8)
    
image = Image.fromarray(image).convert("RGB")
\end{python}

\newpage

\section{Data Format Details of Training}
\subsection{Data Format of Local Label Training Stage}

\begin{tcolorbox}[colback=black!5!white,colframe=black!75!black,title=Local Labels with Grounding]
$<$Img$>$$<$Image feature$>$$<$/Img$>$ [identify] Please describe the critical findings along with their localized bounding boxes in the radiological image of a chest as much detail as possible: \textcolor{blue}{Local diseases of this chest radiograph are $<$p$>$Aortic enlargement$<$/p$>$ $\left\{<53><28><65><36>\right\}$,$<$p$>$Cardiomegaly$<$/p$>$ $\left\{<38><49><82><63>\right\}$.}
\end{tcolorbox}

\subsection{Data Format of Global Label Training Stage}

\begin{tcolorbox}[colback=black!5!white,colframe=black!75!black,title=Global Labels with Diagnosing]
$<$Img$>$$<$Image feature$>$$<$/Img$>$ [vqa] Given the provided chest X-ray image, which of the following diagnoses are present (select all that apply): COPD, Lung Tumor, Pneumonia, Tuberculosis, Other Disease, or No Finding? \textcolor{blue}{Global diseases of this chest radiograph are Lung tumor, Pneumonia, Tuberculosis.}
\end{tcolorbox}
 
\section{Tables for Classification Results}
\label{sec:details_classification_results}
\subsection{Local labels}
The classification results for the critical findings, which include 22 local labels, are presented in Tables \ref{bio_pub_local}, \ref{bio_local}, and \ref{pub_local}, respectively. Each table represent a method we approached. Each table represents a different method.

\begin{table}[!t]
\centering
\begin{tabular}{lcccc}
\hline
\textbf{Local labels}       & \textbf{Precision} & \textbf{Recall} & \textbf{F1-score} & \textbf{Support} \\ \hline
Aortic enlargement & 0.35   & 0.74     & 0.47    & 220  \\
Atelectasis        & 0.00   & 0.00     & 0.00    & 86   \\
Calcification      & 0.50   & 0.01     & 0.01    & 194  \\
Cardiomegaly       & 0.56   & 0.66     & 0.61    & 309  \\
Clavicle fracture  & 0.00   & 0.00     & 0.00    & 2    \\
Consolidation      & 0.00   & 0.00     & 0.00    & 96   \\
Edema              & 0.00   & 0.00     & 0.00    & 0    \\
Emphysema          & 0.00   & 0.00     & 0.00    & 3    \\
Enlarged PA        & 0.00   & 0.00     & 0.00    & 8    \\
ILD                & 0.12   & 0.00     & 0.01    & 221  \\
Infiltration       & 0.33   & 0.02     & 0.03    & 58   \\
Lung Opacity       & 0.12   & 0.07     & 0.09    & 84   \\
Lung cavity        & 0.00   & 0.00     & 0.00    & 9    \\
Lung cyst          & 0.00   & 0.00     & 0.00    & 2    \\
Mediastinal shift  & 0.00   & 0.00     & 0.00    & 20   \\
Nodule/Mass        & 0.31   & 0.02     & 0.04    & 176  \\
Pleural effusion   & 0.24   & 0.57     & 0.34    & 111  \\
Pleural thickening & 0.28   & 0.58     & 0.38    & 169  \\
Pneumothorax       & 0.00   & 0.00     & 0.00    & 18   \\
Pulmonary fibrosis & 0.46   & 0.29     & 0.36    & 217  \\
Rib fracture       & 0.00   & 0.00     & 0.00    & 11   \\
Other lesion       & 0.00   & 0.00     & 0.00    & 94   \\ \hline
micro avg          & 0.36   & 0.29     & 0.32    & 2108 \\
macro avg          & 0.15   & 0.13     & 0.11    & 2108 \\
weighted avg       & 0.30   & 0.29     & 0.23    & 2108 \\
samples avg        & 0.32   & 0.32     & 0.29    & 2108 \\ \hline
\end{tabular}
\caption{Performance metrics for local labels using BiomedCLIP + PubMedCLIP - Llama 2.}
\label{bio_pub_local}
\end{table}

\begin{table}[!t]
\centering
\begin{tabular}{lcccc}
\hline
\textbf{Local labels}       & \textbf{Precision} & \textbf{Recall} & \textbf{F1-score} & \textbf{Support} \\ \hline
Aortic enlargement & 0.31   & 0.59     & 0.41    & 220  \\
Atelectasis        & 0.09   & 0.01     & 0.02    & 86   \\
Calcification      & 0.00   & 0.00     & 0.00    & 194  \\
Cardiomegaly       & 0.56   & 0.58     & 0.57    & 309  \\
Clavicle fracture  & 0.00   & 0.00     & 0.00    & 2    \\
Consolidation      & 0.17   & 0.02     & 0.04    & 96   \\
Edema              & 0.00   & 0.00     & 0.00    & 0    \\
Emphysema          & 0.00   & 0.00     & 0.00    & 3    \\
Enlarged PA        & 0.00   & 0.00     & 0.00    & 8    \\
ILD                & 0.50   & 0.02     & 0.04    & 221  \\
Infiltration       & 0.67   & 0.03     & 0.07    & 58   \\
Lung Opacity       & 0.10   & 0.04     & 0.05    & 84   \\
Lung cavity        & 0.00   & 0.00     & 0.00    & 9    \\
Lung cyst          & 0.00   & 0.00     & 0.00    & 2    \\
Mediastinal shift  & 0.00   & 0.00     & 0.00    & 20   \\
Nodule/Mass        & 0.00   & 0.00     & 0.00    & 176  \\
Pleural effusion   & 0.23   & 0.68     & 0.35    & 111  \\
Pleural thickening & 0.19   & 0.51     & 0.28    & 169  \\
Pneumothorax       & 0.00   & 0.00     & 0.00    & 18   \\
Pulmonary fibrosis & 0.38   & 0.40     & 0.39    & 217  \\
Rib fracture       & 0.00   & 0.00     & 0.00    & 11   \\
Other lesion       & 0.00   & 0.00     & 0.00    & 94   \\ \hline
micro avg          & 0.31   & 0.27     & 0.29    & 2108 \\
macro avg          & 0.15   & 0.13     & 0.10    & 2108 \\
weighted avg       & 0.27   & 0.27     & 0.22    & 2108 \\
samples avg        & 0.30   & 0.30     & 0.27    & 2108 \\ \hline
\end{tabular}
\caption{Performance metrics for local labels using BiomedCLIP - Llama 2.}
\label{bio_local}
\end{table}

\begin{table}[!t]
\centering
\begin{tabular}{lcccc}
\hline
\textbf{Local labels}       & \textbf{Precision} & \textbf{Recall} & \textbf{F1-score} & \textbf{Support} \\ \hline
Aortic enlargement & 0.34   & 0.39     & 0.36    & 220  \\
Atelectasis        & 0.33   & 0.01     & 0.02    & 86   \\
Calcification      & 0.00   & 0.00     & 0.00    & 194  \\
Cardiomegaly       & 0.49   & 0.52     & 0.50    & 309  \\
Clavicle fracture  & 0.00   & 0.00     & 0.00    & 2    \\
Consolidation      & 0.00   & 0.00     & 0.00    & 96   \\
Edema              & 0.00   & 0.00     & 0.00    & 0    \\
Emphysema          & 0.00   & 0.00     & 0.00    & 3    \\
Enlarged PA        & 0.00   & 0.00     & 0.00    & 8    \\
ILD                & 1.00   & 0.00     & 0.01    & 221  \\
Infiltration       & 0.00   & 0.00     & 0.00    & 58   \\
Lung Opacity       & 0.04   & 0.01     & 0.02    & 84   \\
Lung cavity        & 0.00   & 0.00     & 0.00    & 9    \\
Lung cyst          & 0.00   & 0.00     & 0.00    & 2    \\
Mediastinal shift  & 0.00   & 0.00     & 0.00    & 20   \\
Nodule/Mass        & 0.00   & 0.00     & 0.00    & 176  \\
Pleural effusion   & 0.32   & 0.12     & 0.17    & 111  \\
Pleural thickening & 0.20   & 0.57     & 0.30    & 169  \\
Pneumothorax       & 0.00   & 0.00     & 0.00    & 18   \\
Pulmonary fibrosis & 0.22   & 0.16     & 0.18    & 217  \\
Rib fracture       & 0.00   & 0.00     & 0.00    & 11   \\
Other lesion       & 0.36   & 0.09     & 0.14    & 94   \\ \hline
micro avg          & 0.30   & 0.19     & 0.23    & 2108 \\
macro avg          & 0.15   & 0.08     & 0.08    & 2108 \\
weighted avg       & 0.30   & 0.19     & 0.17    & 2108 \\
samples avg        & 0.25   & 0.20     & 0.20    & 2108 \\ \hline
\end{tabular}
\caption{Performance metrics for local labels using PubMedCLIP - Llama 2.}
\label{pub_local}
\end{table}

\subsection{Global labels}
The classification results for the diagnoses, which include 6 global labels, are presented in Tables \ref{bio_pub_global}, \ref{bio_global}, and \ref{pub_global}. Each table represents a different method.

\begin{table}[!t]
\centering
\begin{tabular}{lcccc}
\hline
\textbf{Local labels}       & \textbf{Precision} & \textbf{Recall} & \textbf{F1-score} & \textbf{Support} \\ \hline
COPD          & 0.33 & 0.50 & 0.40 & 2    \\
Lung tumor    & 0.23 & 0.15 & 0.18 & 80   \\
Pneumonia     & 0.56 & 0.40 & 0.47 & 246  \\
Tuberculosis  & 0.63 & 0.18 & 0.28 & 164  \\
Other disease & 0.58 & 0.59 & 0.59 & 657  \\
No finding    & 0.83 & 0.95 & 0.89 & 2051 \\ \hline
micro avg     & 0.76 & 0.77 & 0.76 & 3200 \\
macro avg     & 0.53 & 0.46 & 0.47 & 3200 \\
weighted avg  & 0.74 & 0.77 & 0.74 & 3200 \\
samples avg   & 0.78 & 0.80 & 0.79 & 3200 \\ \hline
\end{tabular}
\caption{Performance metrics for global labels using BiomedCLIP + PubMedCLIP - Llama 2.}
\label{bio_pub_global}
\end{table}

\begin{table}[!t]
\centering
\begin{tabular}{lcccc}
\hline
\textbf{Local labels}       & \textbf{Precision} & \textbf{Recall} & \textbf{F1-score} & \textbf{Support} \\ \hline
COPD          & 0.00 & 0.00 & 0.00 & 2    \\
Lung tumor    & 0.35 & 0.24 & 0.28 & 80   \\
Pneumonia     & 0.59 & 0.34 & 0.43 & 246  \\
Tuberculosis  & 0.51 & 0.23 & 0.32 & 164  \\
Other disease & 0.62 & 0.54 & 0.58 & 657  \\
No finding    & 0.82 & 0.97 & 0.89 & 2051 \\ \hline
micro avg     & 0.76 & 0.77 & 0.77 & 3200 \\
macro avg     & 0.48 & 0.39 & 0.42 & 3200 \\
weighted avg  & 0.73 & 0.77 & 0.74 & 3200 \\
samples avg   & 0.79 & 0.80 & 0.79 & 3200 \\ \hline
\end{tabular}
\caption{Performance metrics for global labels using BiomedCLIP - Llama 2.}
\label{bio_global}
\end{table}

\begin{table}[!t]
\centering
\begin{tabular}{lcccc}
\hline
\textbf{Local labels}       & \textbf{Precision} & \textbf{Recall} & \textbf{F1-score} & \textbf{Support} \\ \hline
COPD          & 0.00 & 0.00 & 0.00 & 2    \\
Lung tumor    & 0.00 & 0.00 & 0.00 & 80   \\
Pneumonia     & 0.37 & 0.27 & 0.31 & 246  \\
Tuberculosis  & 0.34 & 0.20 & 0.25 & 164  \\
Other disease & 0.52 & 0.51 & 0.51 & 657  \\
No finding    & 0.80 & 0.92 & 0.85 & 2051 \\ \hline
micro avg     & 0.70 & 0.72 & 0.71 & 3200 \\
macro avg     & 0.34 & 0.32 & 0.32 & 3200 \\
weighted avg  & 0.66 & 0.72 & 0.69 & 3200 \\
samples avg   & 0.74 & 0.75 & 0.74 & 3200 \\ \hline
\end{tabular}
\caption{Performance metrics for global labels using PubMedCLIP - Llama 2.}
\label{pub_global}
\end{table}

\end{document}